# Three-fold Superstructured Superlattice HfN/HfAlN Thin Films for Enhanced Toughness

Marcus Lorentzon[1,*], Rainer Hahn[2], Justinas Palisaitis[1], Helmut Riedl[2], Lars Hultman[1,3], Jens Birch[1], Naureen Ghafoor[1,*]

[1] Thin Film Physics Division, Department of Physics, Chemistry and Biology (IFM), Linköping University, Sweden

[2] Christian Doppler Laboratory for Surface Engineering of high-performance Components, TU Wien, A-1060 Wien, Austria

[3] Center for Plasma and Thin Film Technologies, Ming Chi University of Technology, 84 Gungjuan Rd., Taishan Dist. New Taipei City 24301, Taiwan

*marcus.lorentzon@liu.se

*naureen.ghafoor@liu.se

# Abstract

To simultaneously achieve high hardness and high toughness in protective coatings remains a fundamental challenge. Here, we harness the superlattice architecture to combine Koehler hardening while the coherent interfaces reduce the crack driving force and improve toughness, enabling coatings that are both hard and damage tolerant. We design and fabricate epitaxial $HfN_{1.33}/Hf_{0.76}Al_{0.24}N_{1.15}$ superlattices, deposited on MgO(001) substrates using low-energy, high-flux ion-assisted reactive magnetron sputtering. These superlattices with bilayer periods ranging from 6 to 20 nm, exhibit a unique three-fold superstructure, confirmed by X-ray diffraction and reciprocal space mapping (RSM). Each constituent forms distinct 3D checkerboard superstructures, with a period of 7.5 Å for $HfN_{1.33}$ and 12.5 Å for $Hf_{0.76}Al_{0.24}N_{1.15}$, that remain clearly resolvable in the combined superlattice architecture. RSMs further reveal low mosaicity, high crystalline quality, and in-plane compressive strains, indicating well preserved coherence across interfaces. Mechanical testing shows that the superlattices maintain the high hardness of $Hf_{0.76}Al_{0.24}N_{1.15}$ (~36 GPa) independent of bilayer period, while surpassing the softer $HfN_{1.33}$ (~27 GPa), consistent with interface-driven Koehler strengthening. Micropillar compression shows brittle fracture on the {110}<110> system, yet with distributed cracking and faster mechanical recovery compared to monolithic films, suggesting improved toughness. Cube-corner indentation further corroborate this behavior, with pile-up and suppressed fracture events. These results demonstrate that epitaxial $HfN_{1.33}$ / $Hf_{0.76}Al_{0.24}N_{1.15}$ superlattices uniquely combine high hardness with improved toughness, enabled by their three-fold superstructured architecture. Leveraging the intrinsic high-temperature stability of HfN-based materials, this design offers a robust pathway toward next-generation protective coatings capable of maintaining performance under extreme conditions.

# 1 Introduction

The multilayer design strategy offers a powerful tool to enhance the fracture toughness, and thus the overall performance, of inherently brittle ceramic hard coatings. By tailoring key parameters such as constituent materials, microstructural architectures, layer thicknesses, and thickness ratios, multilayers enable precise control over mechanical behavior, allowing optimization across a broad spectrum of properties to meet application-specific demands.

Using relatively thick layers, effective crack arrest was demonstrated in a polycrystalline multilayer composed of CrAlN (250 nm) and CrAlSiN (50 nm), with fracture toughness increasing up to 3.2 MPa·√m. This enhancement was attributed to dissimilar fracture pathways, either transgranular or intergranular, in the constituent materials [1]. Similarly, a combination of hard brittle ceramic with a softer, more compliant material can significantly improve fracture toughness through crack deflection mechanisms, as shown in multilayers such as CrN (500 nm) / Cr (250 nm) and TiN (250 nm) / $SiO_x$ (20 nm) [2], [3], [4]. In the CrN/Cr system, the toughness increased to 5.0 MPa·√m compared to 2.7 MPa·√m for Cr and 3.6 MPa·√m for CrN. However, this improvement came at the cost of reduced hardness due to the inclusion of the compliant Cr layers. In such hard-soft material combinations, a substantial reduction in crack driving force occurs when the crack propagates into the softer material [5]. This effect, combined with a high density of interfaces, can lead to a significant increase in fracture toughness [6].

The influence of crystallinity on both hardness and toughness has also been demonstrated in polycrystalline TiN / $Zr_{0.37}Al_{0.63}N_{1.09}$ multilayers with thinner (10 nm / 10 nm) layers. Depending on growth temperature, the ZrAlN layers formed three distinct structures; amorphous, nanocrystalline wurtzite, or decomposed nanolabyrinth structure [7]. These microstructures enabled different toughening mechanisms, with the best combination of properties – high strength and hardness (30 GPa) along with high toughness (2.8 MPa·√m) – was achieved for the nanolabyrinth ZrAlN. This was attributed to effective crack deflection, facilitated by a domain epitaxial relationship with TiN.

Superlattices, a special type of multilayer architecture, are characterized by the epitaxial growth of each layer on top of the preceding one, forming an artificial crystalline structure with a distinct superlattice periodicity ($\Lambda$), in addition to the individual lattice periodicity. This structure gives rise to satellite peaks in X-ray diffraction patterns, the position of which are determined by the bilayer period. More importantly, it has been shown that both hardness and fracture toughness can be simultaneously improved beyond the rule of mixture of the constituent layers, provided an optimal superlattice period is achieved. Examples include TiN/CrN ($\Lambda_{opt} = 6.2$ nm) [8], MoN/TaN ($\Lambda_{opt} = 5.2$ nm) [9], TiN/CrAlN ($\Lambda_{opt} = 7.3$ nm) [10], TiN/WN ($\Lambda_{opt} = 10.2$ nm) [11], and $TiB_2/ZrB_2$ and $TiB_2/WB_2$ ($\Lambda_{opt} \approx 4.0$ nm) [12]. The observed hardness increase is commonly attributed to limited dislocation motion across coherent interfaces [13], a phenomenon known as Koehler hardening [14]. Simultaneous toughness increase is typically associated with mechanisms such as crack deflection or crack arrest at interfaces, driven by elastic or shear modulus mismatch and coherency stress/strain variations arising from differences in lattice parameters.

Inspired by the stark contrast in mechanical behavior in our previous work between overstoichiometric $HfN_y$ [15] and cubic phase (c) HfAlN [16] single-crystals, we attempt to combine these two materials into a single superlattice system aimed at achieving enhanced hardness and toughness. $HfN_y$ exhibits exceptional plasticity and does not fracture under uniaxial compressive loading, owing to high dislocation activity on the {111}<110> slip system, and a relatively low barrier for dislocation nucleation [15]. Despite this ductility, the material retains a high hardness of 26-28 GPa. These properties are attributed to the large overstoichiometry with high densities of point defects (Hf-vacancies and N-interstitials) that self-organize into a 3D checkerboard superstructure. In contrast, c-HfAlN exhibits a significantly higher hardness of ~38 GPa due to inhibited dislocation motion, but shows brittle failure under uniaxial stress, fracturing along the {110}<110> slip system [16]. This behavior is also linked to a 3D checkerboard superstructure, formed of Hf- and Al-rich nanodomains.

In this work, we synthesize single-crystal $HfN_{1.33}$ / $Hf_{0.76}Al_{0.24}N_{1.15}$ superlattices on MgO(001) substrates using ion assisted reactive magnetron sputtering. We investigate the formation of the superlattice structure, the internal 3D superstructures, and the crystalline quality using X-ray diffraction, reciprocal space maps, and scanning transmission electron microscopy. The mechanical properties of the superlattices along with monolithic reference films of $HfN_{1.33}$ and $Hf_{0.76}Al_{0.24}N_{1.15}$ are evaluated using nanoindentation for hardness, micropillar compression for strength and plasticity under uniaxial loading, and cube-corner indentation for qualitative fracture analysis.

# 2 Experimental details

## 2.1 Film Growth and Composition

The $Hf_{1-x}Al_xN_y$ films were deposited in a high vacuum deposition chamber on MgO(001) and Si(001) substrates using two magnetically coupled ϕ = 75 mm (3") unbalanced type II magnetrons. Elemental targets of Hf (99.9% except for a few percent Zr) and Al (99.99 %) were co-sputtered in an Ar + $N_2$ atmosphere at 0.6 Pa (4.5 mTorr) with a $N_2$ partial pressure of 0.067 Pa (0.5 mTorr). The base pressure was better than $1.1*10^{-4}$ Pa ($8.0*10^{-7}$ Torr) at the deposition temperature of 800 °C. Details of the deposition system can be found elsewhere [17], [18], [19]. The MgO substrates were prepared according to the method described in Ref [20]: cleaned by sonication in a detergent solution (Hellmanex III, 2 vol.%), for 5 min, rinsed in de-ionized water followed by 10 min sonication in acetone and ethanol. Finally, the substrates were blow dried with $N_2$ just before inserting into the deposition chamber. An annealing treatment was performed for 1 h at 900 °C prior to deposition.

The magnetron power supplies were operated in constant power mode delivering 125 W to the Hf target (voltage, U = 340 V) and 60 W to the Al target (U = 223 V). Ion assistance was achieved by extending the plasma toward the substrate by magnetically coupling the Hf magnetron field using a stationary electromagnetic coil around the rotating (16 rpm) substrate table. A -30 V substrate bias accelerated low-energy ions to the growing film. Electrical contact was made to the film using a sharp Mo-pin, as the MgO substrate is insulating. As determined in our previous work [16], the ion energy and ion-to-atom flux ratio were $E \approx 20$ eV, and $J_{ion}/J_{atom} \approx 13$ for both $HfN_{1.33}$ and $Hf_{0.76}Al_{0.24}N_{1.15}$. Deposition rates and film compositions were also obtained from monolithic films in ref [16] using time-of-flight elastic recoil detection analysis and Rutherford backscattering spectrometry.

The $HfN_{1.33}/Hf_{0.76}Al_{0.24}N_{1.15}$ superlattices were prepared by periodically blocking the flux of Al atoms by a fast-actuating shutter in front of the Al target. Three multilayer designs of total thickness of ~1 µm, were chosen. The thickest bilayer period was $\Lambda = 20$ nm, i.e. with nominally 10 nm / 10 nm thick $HfN_{1.33}$ / $Hf_{0.76}Al_{0.24}N_{1.15}$ and a total of 50 bilayers. Then a 5 nm / 5 nm design with 100 bilayers and lastly the thinnest design of 3 nm / 3 nm and 167 bilayers.

## 2.2 Structural Characterization

X-ray diffraction (XRD) θ-2θ scans were carried out in a Panalytical X'Pert Pro in Bragg-Brentano geometry with 1/2° slits and monochromatic Cu kα radiation by using a Ni-filter. Reciprocal space maps (RSM) were acquired in a Panalytical Empyrean with high-resolution primary optics using hybrid Cu-Kα monochromator, and open secondary optics with the pixel detector operating in 2D line scan mode for fast RSM acquisition, yet with reasonably high resolution. The RSMs were used to evaluate the superlattice quality, the formation of 3D superstructures, relaxed lattice parameter, limited lateral size, L, (lateral x-ray coherence length) and mosaic tilt. The diffractometer angles were converted into reciprocal space according to equations 2.1 and 2.2 [21].

$$Q_\perp = \frac{4\pi}{\lambda} \sin\theta \cos(\theta - \omega). \qquad 2.1$$

$$Q_\parallel = \frac{4\pi}{\lambda} \sin\theta \sin(\theta - \omega). \qquad 2.2$$

Transmission electron microscopy (TEM) samples were prepared using two methods. A conventional sandwich method by gluing two pieces of the sample together in a TEM grid with the film facing each other, followed by mechanical polishing to ~50 μm thickness and thinning to electron transparency by $Ar^+$ ion milling with 5 keV at 5°. The final step of ion polishing using 1.5 keV for 20 min was used to remove amorphized surface damage. Focused ion beam (FIB) was used to extract electron transparent lamellas from selected compressed micropillars by the conventional lift-out method using a Thermo Scientific Helios 5 UC DualBeam system. Before lift-out and thinning, an electron beam-induced Pt layer was deposited to provide protection from $Ga^+$ implantation. For final thinning, a low-energy milling at 2 kV was performed to minimize Ga-induced damage.

Lattice resolved high-angle annular dark field scanning TEM (HAADF STEM) micrographs were obtained in a double Cs corrected Titan$^3$ 60-300 microscope, operated at 300 kV. Samples were aligned to the MgO(001) zone axis for imaging. Compressed micropillars were imaged by scanning electron microscopy (SEM) at a ~50° tilt angle by a Zeiss Sigma 300 operated at a 3 kV acceleration voltage and using SE2 detector.

## 2.3   Mechanical Characterization

Hardness was measured by nanoindentation in a Triboindentor, Hysitron TI950, equipped with a Berkovich diamond tip. The area function was calibrated by indents in a fused quartz standard. The load-displacement curves were analyzed by the Oliver and Pharr method [22], and 25 indents were made for each sample.

Micropillars from the superlattices and monolithic reference films were fabricated using a ThermoFisher Scios 2 focused ion beam (FIB) – scanning electron microscope (SEM) system via focused ion beam milling. Each pillar had a diameter of approximately 300 nm and an aspect ratio of 3:1. The top diameter and height of each pillar were determined from SEM images before testing to ensure accurate stress and strain calculations. The fabrication process employed a $Ga^+$ ion beam, with a probe current of 7 nA for coarse milling and subsequently reduced to 50 pA for final polishing. Particular care was taken to terminate the milling precisely at the coating–substrate interface to ensure consistent boundary conditions. The final taper angle was below 2° for all pillars. At least 7 pillars per sample were produced and tested to ensure reproducibility.

In-situ pillar compression was done in a Zeiss Sigma 500 VP system (operated in high-vacuum) using FemtoTools FT-NMT04 in-situ SEM nanoindenter equipped with a 7 µm diameter flat diamond punch in a displacement-controlled mode with a displacement rate of 10 nm·s$^{-1}$, corresponding to a nominal strain rate of ~10$^{-2}$ s$^{-1}$. The pillars were loaded to different strains, up to a maximum of ~50% strain. The correct alignment between the sample and tip was guaranteed by the following procedure: Before the actual loading of the pillar, indentation tests were carried out in the pristine coating, making sure that there was a sharp transition between the force recording in the vacuum and the loading of the film. Given that the angle between the two parts cannot be modified with this instrument, and the specimen must always be positioned the same way on the specimen holder due to the instrument's design, a one-time FIB alignment procedure was performed on the flat punch tip.

The load-displacement curves were converted into engineering stress-strain curves following an approach by Wheeler and Michler [23], where the top diameter of the pillar is taken as the reference contact area. The engineering strain was obtained from the displacement data using the pillar height and a corrected displacement data accounting for the base compliance following Sneddon's correction [24] given by equation 2.3.

$$\Delta L = \frac{1 - v_{sub}^2}{E_{sub}} * \frac{F}{d}, \qquad 2.3$$

where $v_{Sub}$ and $E_{Sub}$ are the biaxial Poisson's ratio and Young's modulus of the substrate, respectively, $v_{sub}$ = 0.18, $E_{sub}$ = 291 GPa [25]. $F$ is the applied force, and $d$ is the diameter of the pillars. $\Delta L$ is then the deformation of the substrate induced by the pillar and must be subtracted before calculating the strain.

The yield strength was extracted using a 0.2% offset criterion, determined by constructing a line parallel to the initial linear-elastic slope of the stress–strain curve and identifying its intersection with the experimental curve.

Qualitative fracture toughness analysis was performed using a Micro Materials NanoTest Vantage indenter and a cube-corner tip. The depth-load curves of total of 25 indents per sample was recorded, with increasing maximal load from 10 mN to 100 mN which resulted in depths from ~250 nm to ~1400 nm. The indentation imprints as well as the compressed micropillars were imaged using a Zeiss Sigma 300 scanning electron microscope using a 3 kV acceleration voltage. An InLens detector was used for imaging of the cube-corner imprints and an SE2 detector for imaging of the compressed pillars.

# 3 Results and Discussion

## 3.1 Structural analysis

The structural analysis of the superlattices is supported by our previous investigations of the monolithic films of the constituent layer materials, namely $HfN_{1.33}$ [15] and $Hf_{0.76}Al_{0.24}N_{1.15}$ [16]. Figure 1a shows a fraction of XRD θ-2θ scans (20° - 55°) of these monolithic reference films and three superlattices of varying periodicities.

The diffraction, combined with STEM analysis, reveals that a single-crystal quality $Hf_{0.76}Al_{0.24}N_{1.15}$ film (blue) is obtained on the MgO(100) substrate, exhibiting cube-on-cube epitaxy. An intense 200 peak is centered at 2θ ≈ 39.8° and no other peaks attributed to the c-Hf(Al)N phase is observed in the 2θ range 20°-100°, apart from the symmetric 400 reflection. In addition to the main peak, a broad satellite peak ($S_{HfAlN}$) appears with weak intensity, centered at 2θ = 32.5°. This satellite peak stems from a three-dimensional (3D) chemical modulation within the otherwise single-crystalline material, forming a checkerboard-like pattern aligned along the <100> directions [16], see Figure 1b. The 3D checkerboard superstructure develops during growth near the film surface as a result of spinodal decomposition of the metastable c-HfAlN, activated by energetic Ar-neutrals backscattered from the Hf-target during sputtering. As a consequence, nanometer-sized periodic domains form, enriched in either HfN or AlN, with a periodicity, $\Lambda_z$, of about 12.5 Å for the composition of $Hf_{0.76}Al_{0.24}N_{1.15}$. Another 3D checkerboard superstructure develops in $HfN_{1.33}$ (green) as a broad and weak satellite peak ($S_{HfN}$) appears, centered at 2θ = 28°. In this case, the high overstoichiometry makes $HfN_{1.33}$ form quasi-stoichiometric and hyper-overstoichiometric domains in a checkerboard pattern with a smaller periodicity of ~7.5 Å [15]. These nanoscale domains result from the self-organization of Hf-vacancies and N-interstitials (including N antisite defects), which cluster to minimize the total energy of the material.

When these two materials are combined into a multilayer structure as illustrated in Figure 1c, a three-fold superstructured superlattice emerges. In addition to the underlying single-crystal lattice, three distinct chemical periodicities are present: the 3D checkerboard modulations within $HfN_{1.33}$ and $Hf_{0.76}Al_{0.24}N_{1.15}$, and the periodicity introduced by the multilayering itself. All three contributes to the diffraction pattern in Figure 1a, producing two weak and broad superstructure satellites, along with sharper, more intense superlattice satellites from the multilayer periodicity.

It is worth highlighting that the position of the checkerboard satellites is independent of the superlattice period, remaining fixed at the same angles as for the monolithic films previously reported [15], [16]. In contrast, the superlattice satellites scale with the period, becoming more widely spaced for smaller periods. Using Equation 3.1 [26], [27], the superlattice period is calculated, where $\theta_0$ is the angle of the main peak, $\theta_n$ is the angle of the $n^{th}$ satellite, and n is the order of the satellite, marked in Figure 1a. The obtained superlattice period is slightly smaller than the design, yielding 18.3 nm, 9.2 nm, and 5.4 nm, respectively. The same equation was applied to calculate the period of the 3D superstructure.

$$\Lambda = \frac{n\lambda}{2|\sin\theta_0 - \sin\theta_n|}. \qquad 3.1$$

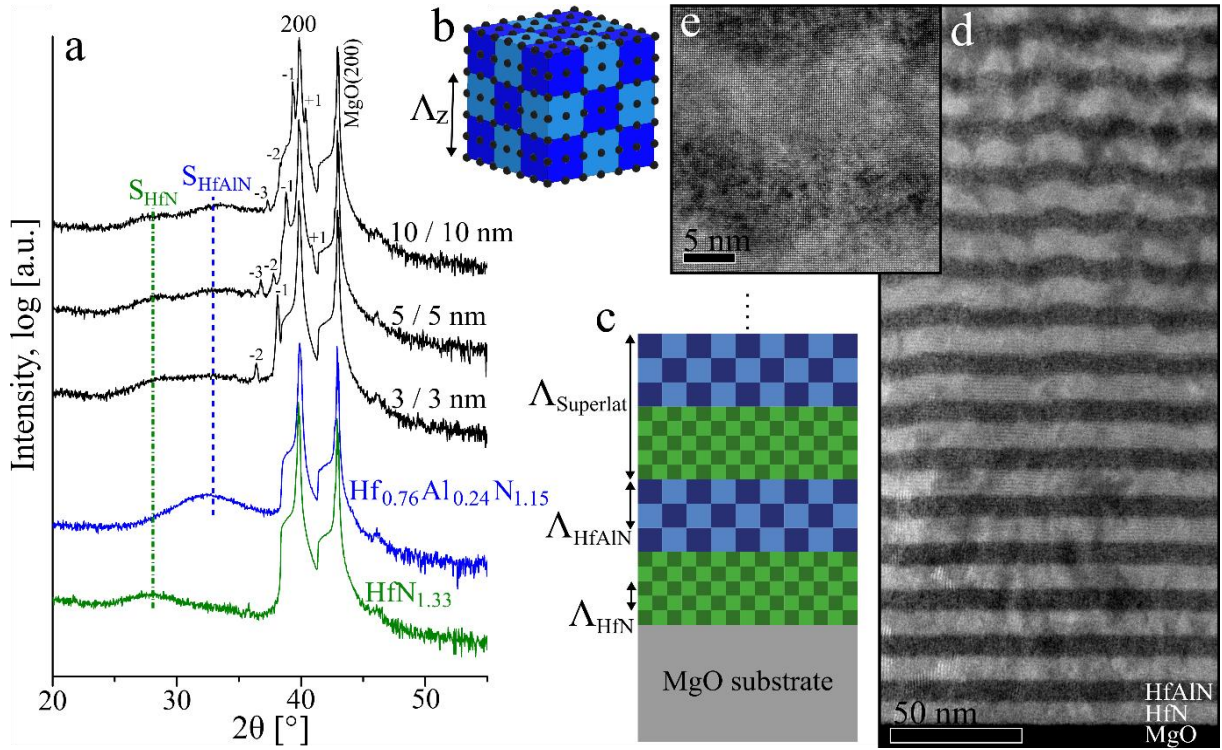

Figure 1: a) XRD θ-2θ scans between 20-55° (vertically offset), intensity in log-scale. The green and blue curves are from monolithic $HfN_{1.33}$ and $Hf_{0.76}Al_{0.24}N_{1.15}$ films, respectively, and superlattices in black. 3D checkerboard satellites are marked "$S_{HfN}$" and "$S_{HfAlN}$", and superlattice satellites with the order number. b) illustrates the 3D checkerboard superstructure as domains of different chemical composition in the lattice. c) illustrates the superlattice architecture and the three different superstructure periods. d-e) show cross-sectional HAADF STEM images along the [001] zone axis of the 10 nm / 10 nm superlattice, e) is taken from the top part of the film.

HAADF STEM imaging of the 10 / 10 nm superlattice in Figure 1d-e show smooth and well-defined epitaxial layers for the first ~160 nm (=8 bilayers). Beyond, accumulated roughness causes a transition into more misoriented layers, yet the epitaxy is maintained. After ~300 nm, the roughness saturates and remains constant up to the final film thickness of ~1 µm. The constant roughness after some depth can be explained by strain relaxation causing local lattice distortion at the surface, resulting in a Stranski-Krastanov growth mode [28], [29]. Despite the roughness, x-ray diffraction shows distinct superlattice peaks as the layer thickness ratio is preserved.

The speckled contrast in both the bright $HfN_{1.33}$ layers and dark $Hf_{0.76}Al_{0.24}N_{1.15}$ layers evidences the 3D checkerboard superstructure [15], [16]. Relatively narrow columnar domains spanning over multiple layers are visible by the contrast variations. Diffraction contrast in the otherwise Z-contrast imaging highlights these domains, which appear brighter when perfectly aligned with the zone axis.

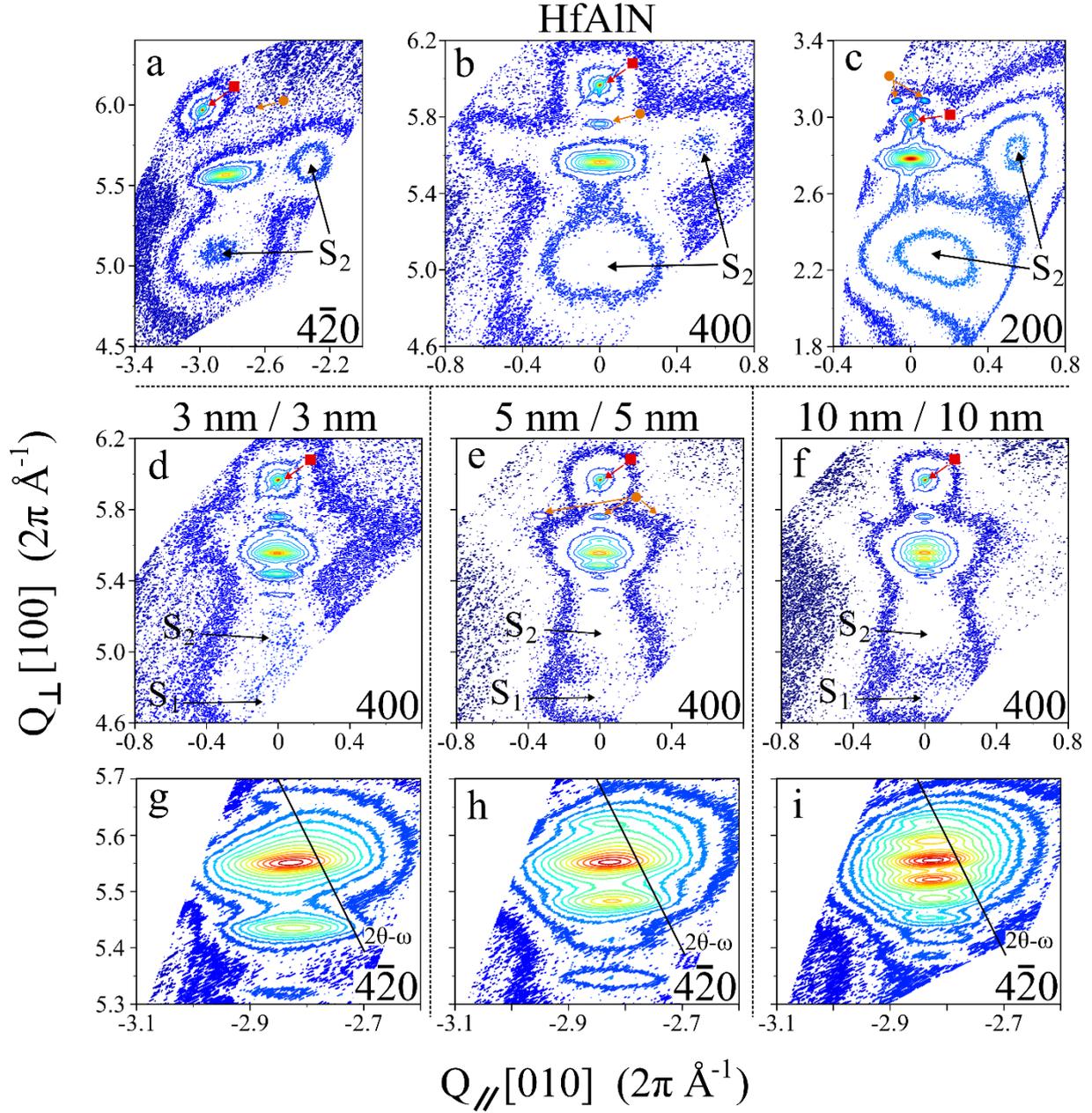

Figure 2: Reciprocal space maps (RSMs) of the reference $Hf_{0.76}Al_{0.24}N_{1.15}$ film and the superlattices with the x-axis and y-axis aligned with the [010] and [100] directions, respectively. The intensity is plotted in log-scale to highlight the weak satellites. a-c) shows RSMs of $4\bar{2}0$, 400, and 200 spots of $Hf_{0.76}Al_{0.24}N_{1.15}$, where the clearly visible 3D superstructure satellites are indicated with "$S_2$". d-f) shows the RSMs of respective superlattice around the symmetric 400 spot, with "$S_2$" indicating the $Hf_{0.76}Al_{0.24}N_{1.15}$ superstructure, and "$S_1$" indicating the $HfN_{1.33}$ superstructure. g-i) shows the RSMs of respective superlattice around the asymmetric $4\bar{2}0$ spot where the superlattice satellites can be clearly distinguished. The solid line in g-i) indicates the $2\theta\text{-}\omega$ line from origin through MgO($4\bar{2}0$) reciprocal spot. The red squares in a-f) indicates the spot from the single crystal MgO substrate. The orange circles in a-c) and e) indicate spots from an unidentified minority phase.

All superlattices exhibit a high crystalline quality, evident by the separation of Cu K$\alpha_1$ and K$\alpha_2$ peaks in XRD, in both the main 200 as well as satellite peaks, closely resembling the peak shape of single-crystal MgO. To further assess crystal quality, lattice parameter, strain, mosaicity, and the presence of superstructures, symmetric and asymmetric reciprocal space maps (RSMs) were recorded around $4\bar{2}0$, 400, and 200 reflections for all superlattices and monolithic reference films.

Figure 2a-c presents wide area RSMs for the monolithic $Hf_{0.76}Al_{0.24}N_{1.15}$ film with the x- and y-axis aligned along the [010] and [100] directions, respectively. Both the c-$Hf_{0.76}Al_{0.24}N_{1.15}$ and the MgO substrate reflections are captured within the same maps. Multiple 3D superstructure satellites labeled $S_2$ are clearly resolved. The intensity of the satellites is asymmetric around the main spot, more intense towards the origin, consistent with the XRD profile in Figure 1 and selected area electron diffraction as previously reported [16]. Additionally, some weak reflections, highlighted by orange circles, appear at positions inconsistent with either the cubic or hexagonal wurtzite phase of HfAlN. These features are attributed to small amounts of a hitherto unidentified phase, which is discussed below. The same phase is also observed in the superlattices and the $HfN_{1.33}$ film, see supplementary information section 1.

The wide-area RSMs of the superlattices in Figure 2d-f around the symmetric 400 peak show multiple relatively intense superlattice satellites that are vertically aligned and equidistantly spaced from each other with a distance that scale inversely with the superlattice period. The maps further confirm the presence of two distinct 3D superstructures, one from $Hf_{0.76}Al_{0.24}N_{1.15}$ and one from $HfN_{1.33}$. However, no clear satellite appears along the lateral directions, suggesting reduced checkerboard quality.

The satellites are better observed in higher-resolution scans in Figure 2g-i, which were used to extract the average lattice parameter, strain, mosaicity, and related values, see Table 1. The RSMs around the asymmetric $4\bar{2}0$ spot show the main reflection along with vertically aligned and equidistant superlattice satellites. All spots are elliptical-like and display pronounced lateral elongation compared to the transverse direction, indicating a large transverse, but relatively small lateral characteristic coherence length. The lateral broadening of the spot is due to relatively narrow columnar grains or a large density of vertical threading dislocations, both of which interrupt the lateral x-ray coherence [30]. The small tilt of the elliptically shaped off-specular diffraction spots is a sign that the superlattices (and reference films) have only a quite small degree of mosaic spread. In cases where the broadening is dominated by mosaic spread, the spot broadens in the ω-direction, i.e. perpendicular to the 2θ-ω line in the figures, rather than the lateral direction. Together with STEM results in Figure 1, we conclude that the films are composed of relatively narrow columnar grains in the growth direction, with only a very small degree of mosaicity with respect to each other. The calculations for limited lateral size and mosaic spread are shown in supplementary information section 1.

Table 1: Calculated values of relaxed lattice parameter, lateral strain, limited lateral size and mosaic spread using the reciprocal space maps.

|  | Lattice param. $a_0$ [Å] | Lateral strain [%] | Limited lateral size [Å] | Mosaic spread [°] |
| --- | --- | --- | --- | --- |
| $HfN_{1.33}$ | 4.506 | -0.855 | 102.8 | 0.014 |
| $Hf_{0.76}Al_{0.24}N_{1.15}$ | 4.478 | -1.268 | 149.5 | 0.179 |
| 3 nm / 3 nm | 4.493 | -1.128 | 110.5 | 0.065 |
| 5 nm / 5 nm | 4.493 | -1.108 | 108.8 | 0.060 |
| 10 nm / 10 nm | 4.492 | -1.043 | 117.0 | 0.069 |

In the case of a fully relaxed lattice, the position of the diffraction spots corresponding to the average lattice spacings of the superlattices should lie on the 2θ-ω line between origin and the MgO($4\bar{2}0$) spot in Figure 2g-i. This is not the case here as the spot clearly deviates from the relaxed position, indicating a relatively large strain in the films, such that the lattice constant in lateral and transverse directions deviates. The relaxed lattice parameter can be calculated using the asymmetric $4\bar{2}0$ spot and equation 3.2 [31] (assuming a Poisson ratio, $\nu$, of 0.25 [32]) where the lateral and transverse lattice parameters are directly calculated from the spot position in reciprocal space by $a_\parallel = \frac{2\pi}{q_\parallel}\sqrt{h^2 + k^2}$ and $a_\perp = \frac{2\pi}{q_\perp}\sqrt{l^2}$, respectively.

The relaxed lattice parameter, shown in Table 1, decreases upon addition of Al, from 4.506 Å for $HfN_{1.33}$ to 4.478 Å for $Hf_{0.76}Al_{0.24}N_{1.15}$ due to incorporation of the smaller Al-atom in the cubic lattice, and agrees well with previous reports [33], [34] for the compounds of concern, albeit on the small side compared to our previous work [15]. The lattice parameter of all three superlattices is more or less constant at $a_0$ = 4.492 Å, just in between the values of the two constituent materials, i.e. the average of the lattice parameters considering the near equal layer thicknesses.

$$a_0 = a_\perp \left(1 - \frac{2\nu(a_\perp - a_\parallel)}{a_\parallel(1 + \nu)}\right). \quad 3.2$$

A lateral compressive strain of -0.855% for $HfN_{1.33}$ and -1.268% for $Hf_{0.76}Al_{0.24}N_{1.15}$ was obtained using equation 3.3, that matches very well with previously reports on epitaxial $Hf_{1-x}Al_xN_y$ films of similar composition [33], [34]. The superlattice strain is also in between the constituent materials, between -1.04 and 1.13%, quite close to the strain in $Hf_{0.76}Al_{0.24}N_{1.15}$.

$$\varepsilon = \frac{a_\parallel - a_0}{a_0}. \quad 3.3$$

## 3.2 Unidentified phase

RSM analysis reveals small amounts of an additional minority phase, evident from the weak, narrow reflections present in all films (marked in Figure 2a-c, e). Known Hf-N phases in the ICDD database were examined and the search expanded to include spinel phases based on Hf, Mg, O, and N, since such phases may form during high temperature growth of transition metal nitrides on MgO substrates [35]. Surface oxides are also expected upon air exposure. However, the unidentified phase could not be matched to any entry. Nor was it detected by STEM, giving no clues to its location within the film. A previous study of epitaxial $Hf_{1-x}Al_xN$ thin films ($0 \leq x \leq 0.71$) showed a similar unidentified peak near the HfN(400) peak, observed in all cubic films, but absent in hexagonal films. The peak position remained fixed at the $2\theta$ angle of ~90° up to $x = 0.3$, while the main c-HfAlN peaks shifted to higher angles, suggesting that the unidentified phase forms in the cubic compounds and is unaffected by Al concentration [16].

The small full width at half maximum (FWHM) of the unidentified peaks and their similar shape (elongated in the lateral direction) with respect to the c-HfAlN peaks, suggests formation of relatively large columnar grains of the unidentified phase. The strong correlation between the unidentified phase and the c-HfAlN was investigated by ɸ-scans in Figure 3 using the same high-resolution XRD setup as for RSM, revealing that the unidentified peak near 200 (Figure 2c) aligns with the c-HfN($4\bar{2}0$) reference. Likewise, the unidentified peak near $4\bar{2}0$ (Figure 2a) aligns only with c-HfN($4\bar{2}0$) (not shown here). Taken together, the constant peak positions, small FWHM, strong correlation to c-HfN and symmetry consideration suggest that the unidentified phase contains Hf and N, possibly also Mg or O, and exhibits four-fold symmetry, consistent with a cubic or tetragonal structure. The plane spacings for the peaks were calculated, see supplementary information section 2. They indicate a relatively complex lattice with multiple tightly spaced planes, yet a cubic symmetry from the strong correlation with the cubic rocksalt HfN peaks.

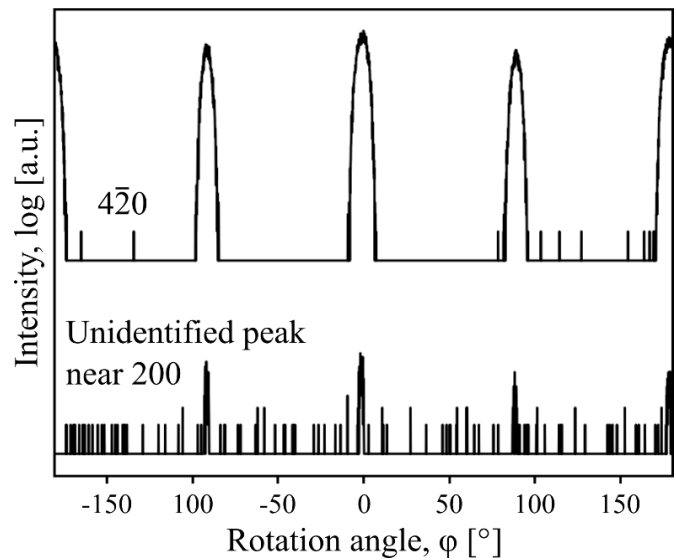

Figure 3: ɸ-scan of HfN($4\bar{2}0$) and the unidentified peaks near 200 in Figure 2c, showing a strong correlation.

## 3.3  Mechanical properties

Figure 4a shows the nanoindentation hardness of monolithic films and superlattices as a function of bilayer period. Although approximately half of the superlattices are composed of layers of the softer HfN$_{1.33}$ ($H = 27.7 \pm 0.9$ GPa), the hardness remains high, near the value of Hf$_{0.76}$Al$_{0.24}$N$_{1.15}$ ($H = 37.0 \pm 0.8$ GPa). A slight decrease can be seen as the bilayer period increases, from $H = 36.4 \pm 1.0$ GPa for $\Lambda = 6$ nm, to $H = 35.6 \pm 1.4$ GPa for $\Lambda = 20$ nm. This behavior is explained by an effective superlattice hardening [14], where dislocation activity in the HfN$_{1.33}$ layers are blocked by the coherent interfaces to Hf$_{0.76}$Al$_{0.24}$N$_{1.15}$. The high hardness of Hf$_{0.76}$Al$_{0.24}$N$_{1.15}$ is in turn explained by the spinodally decomposed structure, where coherency strains between the HfN-rich and AlN-rich domains block dislocation motion [16]. The superlattices thus obtain a similar hardness as Hf$_{0.76}$Al$_{0.24}$N$_{1.15}$.

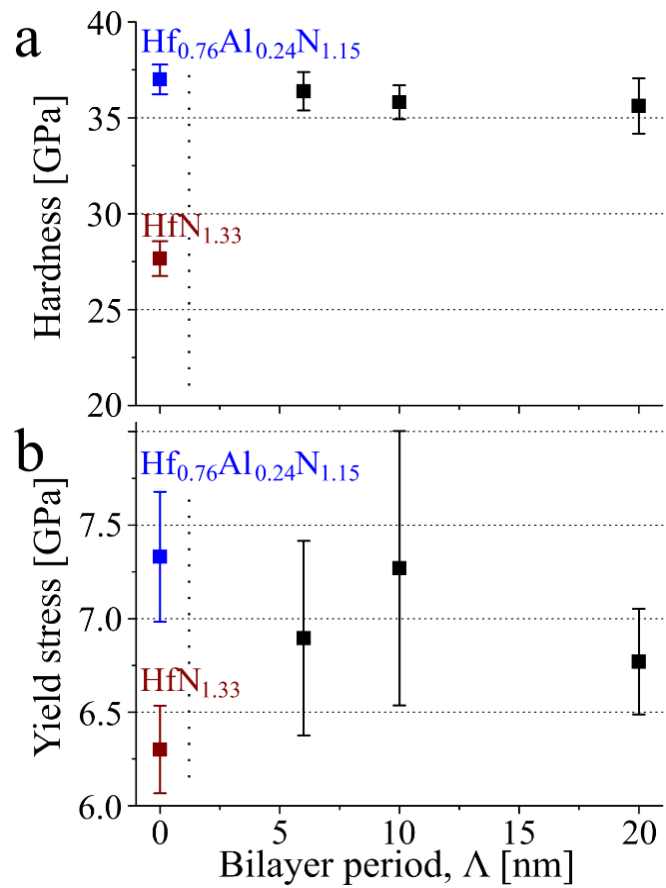

Figure 4: a) Nanoindentation hardness of the monolithic HfN$_{1.33}$ (red) and Hf$_{0.76}$Al$_{0.24}$N$_{1.15}$ (blue) films, and of the superlattices (black) as a function of the bilayer period, $\Lambda$. b) Corresponding yield stress extracted from the stress-strain data of micropillar compression testing.

In contrast, a substantial increase in hardness with respect to monolithic films of both materials in the superlattice is often observed, where the maximum hardness peaks at relatively small superlattice periods of a few nanometers. This phenomenon was shown for example for TiN/VN superlattices ($H > 50$ GPa at $\Lambda \approx 5$ nm) [13], metal/nitride W/NbN superlattices ($H \approx 33$ GPa at $\Lambda = 2\text{-}3$ nm) [36] and TiN/WN superlattices ($H \approx$ 37 GPa at $\Lambda \approx 8$ nm) [11]. Common for all these materials is that dislocation activity is not prohibited, such that the superlattice effects operate in both constituent materials, whereas in this work, the $Hf_{0.76}Al_{0.24}N_{1.15}$ does not allow dislocation motion, thus acting as an upper limit for hardness.

Next, micropillar compression tests were performed to evaluate the strength and plasticity of the superlattices, compared to the reference monolithic films. In contrast to hardness testing, which provides an averaged measure of resistance to plastic deformation, stress–strain curves from micropillar compression reveal the underlying deformation pathways and fracture events. Figure 4b shows the yield strength as a function of the bilayer period, where the superlattices exhibit a high strength of ~7.0 GPa, slightly above the rule-of-mixing between the reference $HfN_{1.33}$ and $Hf_{0.76}Al_{0.24}N_{1.15}$ strengths (6.3 $\pm$ 0.2 GPa and 7.3 $\pm$ 0.3 GPa, respectively), in contrast to hardness.

The engineering stress-strain curves of the monolithic films are shown in Figure 5a where two completely different behaviors are obtained. $HfN_{1.33}$ exhibits a large degree of dislocation-mediated plasticity, first reported by our group [15], attributed to the large amounts of Hf-vacancies and N-interstitials in the checkerboard superstructure. A linear strain hardening is observed up to very large strains of more than 30 % as the pillar deforms plastically, see Figure 5b. In contrast, $Hf_{0.76}Al_{0.24}N_{1.15}$ exhibits the more expected (for transition metal nitrides) brittle fracture upon loading above ~2.5 % strain, identified as a sharp drop in stress. Upon further loading, the stress continues to build until another fracture event. At ~10-12 % strain, a large fracture occurs, where the stress first drops substantially followed by a long period of continually reducing stress as the strain is increased. This fracture is identified as the large crack event at the bottom of the pillar in Figure 5c, while the smaller cracks are responsible for the smaller drops in the stress-strain curves. The overall deformation behavior was consistently observed across all tested pillars, as reflected by the scatter shown in Figure 4b and the representative curves in Figure 5a. While the sequence of individual fracture events varied, yield strength and fracture mode trends were reproducible.

The results from the monolithic films show potential for improved properties by combining the materials in a superlattice, where we propose to combine the high hardness and yield strength of $Hf_{0.76}Al_{0.24}N_{1.15}$ with the great plasticity of $HfN_{1.33}$. Figure 5d-i shows the engineering stress-strain curves and SEM micrographs of the three superlattices. All exhibit brittle fracture in a very similar way to the $Hf_{0.76}Al_{0.24}N_{1.15}$ reference. However, the severity of the fracture events is reduced, as the drop in stress is smaller and the "recovery" is faster. This recovery reflects the ability of the interfaces to arrest or deflect cracks, preventing catastrophic propagation and allowing the pillar to sustain further loading. This is clearly observed for the 3 nm / 3 nm superlattice in Figure 5d, where only minor fracture events is observed up to ~20 % strain. On the other hand, many more fracture events occur compared to $Hf_{0.76}Al_{0.24}N_{1.15}$, confirmed by SEM of the pillars in Figure 5e. The superlattice with thickest layers, 10 nm / 10 nm in Figure 5h-i, exhibit fewer but larger fracture events, compared to the 3 nm / 3 nm superlattice. Still, they are quickly "recovered". Here, the lower interface density reduces the number of effective barriers for crack deflection, so fewer but larger cracks form. Nevertheless, the remaining interfaces still limit catastrophic propagation, allowing partial stress recovery after fracture. Unlike the monolithic coating, which typically failed along one or two dominant shear planes (Figure 5c), the superlattices displayed multiple shear planes distributed along the pillar height. This fragmentation of shear localization indicates that interfaces act as barriers and deflectors, promoting a more distributed fracture pattern. Taken together, the pillar compression tests demonstrate a systematic evolution of the fracture mode with bilayer period: higher interface density fragments cracks into many smaller shear planes, whereas lower interface density allows fewer but larger cracks to form. Thus, strong indications of an improved fracture toughness is obtained at a higher interface density.

Notably, fracture toughness tests on superlattices typically exhibit a maximum $K_{IC}$ at certain bilayer periods, reflecting resistance to crack initiation [8], [9]. However, such tests end once a crack forms. In contrast, micropillar experiments capture the mechanical response after crack initiation, allowing us to assess how cracks propagate and interact with interfaces. This post-initiation behavior explains why we observe a systematic evolution of fracture mode with bilayer period rather than a simple peak in toughness.

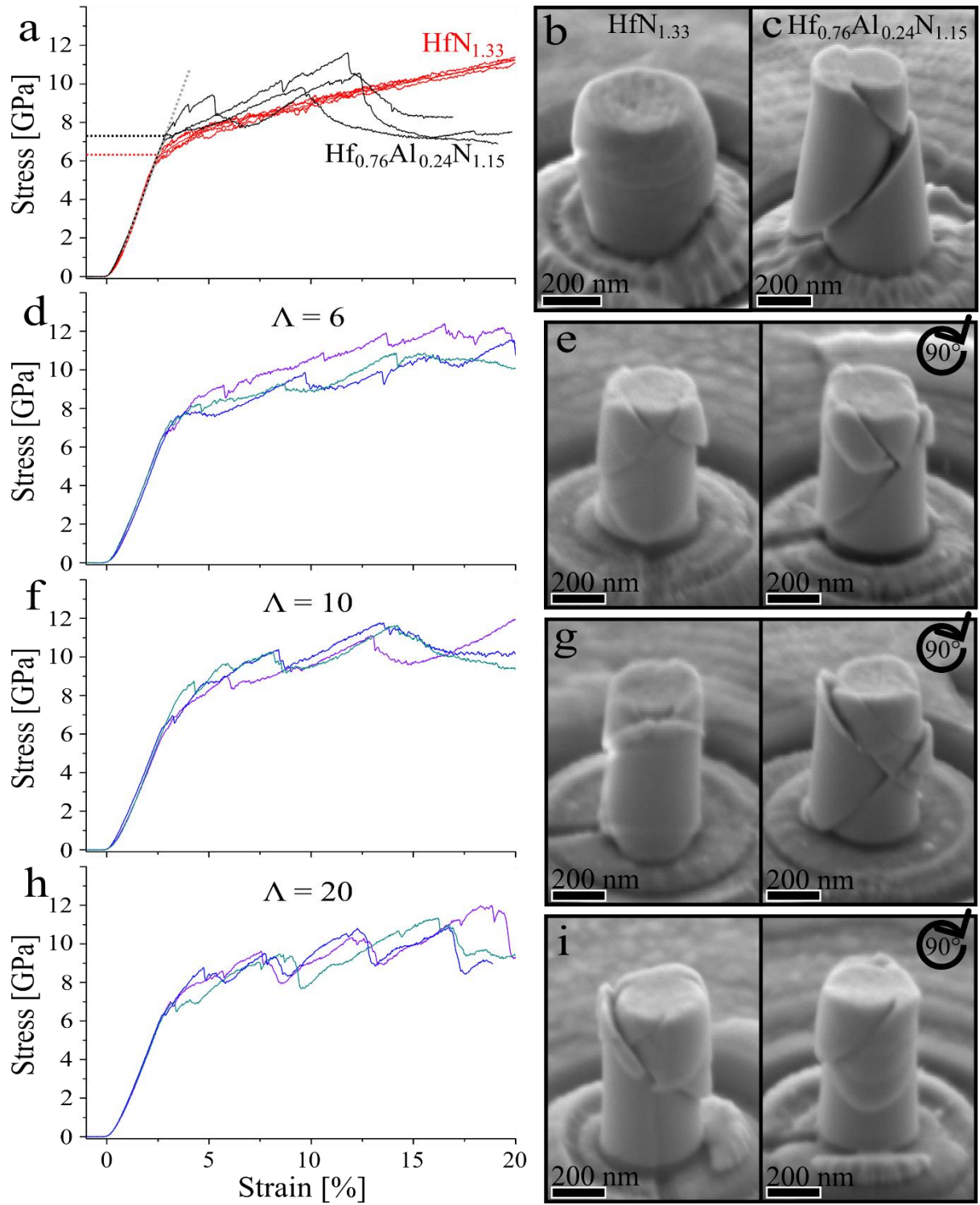

Figure 5: Representative engineering stress-strain curves and SEM micrographs of the compressed pillars at a tilt angle of ~52° with respect to the sample normal. a-c) Reference monolithic $HfN_{1.33}$ and $Hf_{0.76}Al_{0.24}N_{1.15}$ films, respectively. The dotted lines in a) indicate the yield stresses. Note the integrity of plastically deforming $HfN_{1.33}$ and the characteristic fracture slip of $Hf_{0.76}Al_{0.24}N_{1.15}$. d-i) Stress-strain curves and SEM micrographs (viewed in two orientations, 90° rotation) for superlattice pillars: d-e) $\Lambda = 6$ nm, f-g) $\Lambda = 10$ nm, and h-i) $\Lambda = 20$ nm.

An electron transparent lamella of the compressed pillar in Figure 5i was extracted using FIB lift-out technique. The sample was oriented along the <001> axis to allow for lattice resolved imaging. Figure 6a shows a cross-sectional HAADF STEM overview of the $\Lambda = 20$ nm pillar in the [001] zone axis. Six fractures can be identified, where one is a "double" fracture (Figure 6b), all of which have a 45° tilt with respect to the substrate normal. Since the sample was oriented in the [001] zone axis we can easily identify the fractures on the {110}<110> slip system, common with c-HfAlN [16]. The straight and flat fractures, with no deviations in the HfN layers, shows that crack deflection does not occur, instead the toughness enhancement may stem from the crack fragmentation.

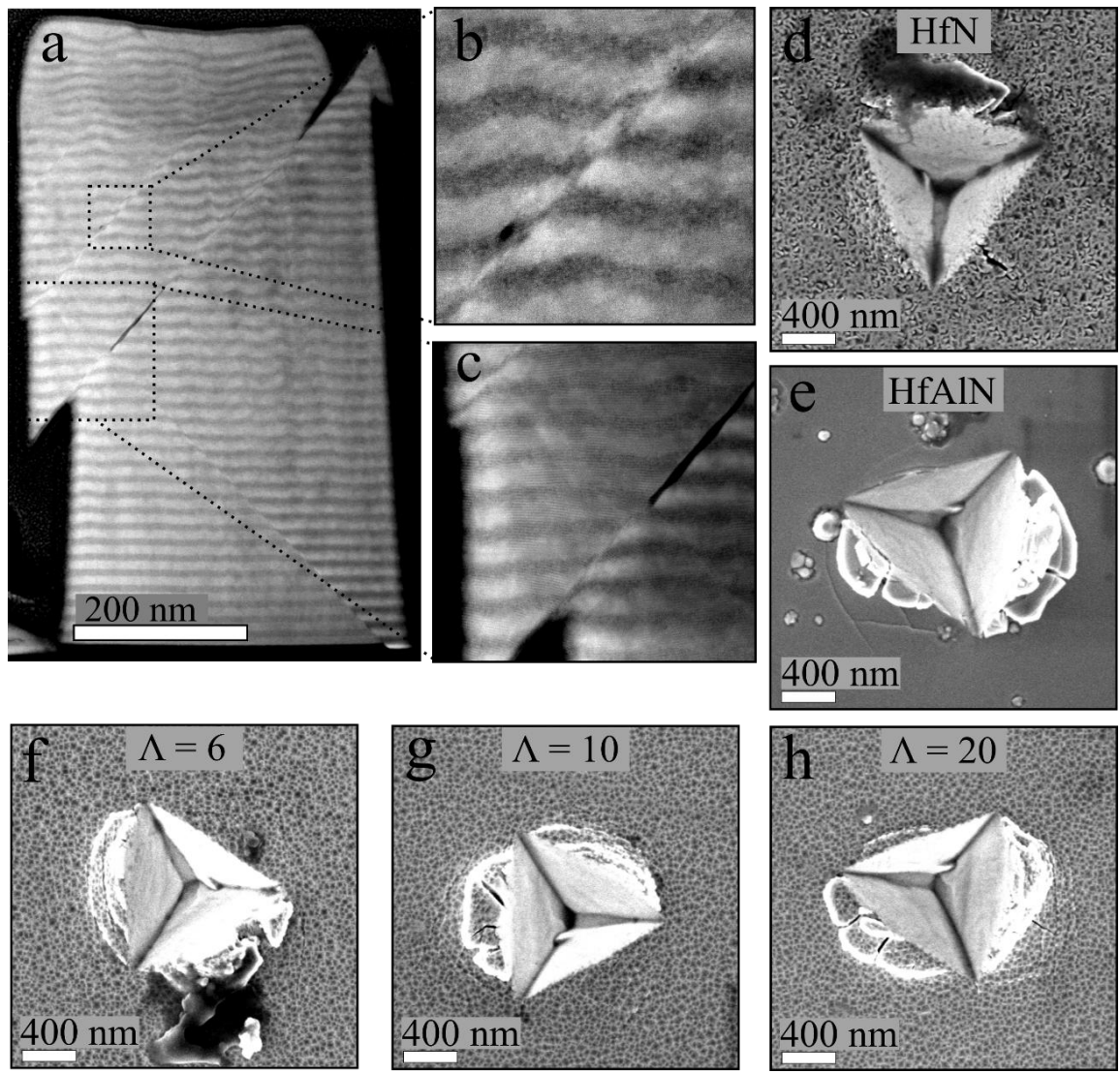

Figure 6: a-c) cross-sectional HAADF-STEM micrographs from the compressed pillar of the superlattice with $\Lambda = 20$ nm in Figure 5f. d-h) Top-view SEM micrographs of the imprints from cube-corner indents at a load of 25 mN, resulting in a depth of about 500 nm, i.e. half the film thickness, d-e) from the reference $HfN_{1.33}$ and $Hf_{0.76}Al_{0.24}N_{1.15}$ monolithic films and f-h) from the superlattices.

The toughness of the films was further assessed using cube-corner indentation, with the resulting imprints analyzed by SEM micrographs (Figure 6d-h) of the indents at a load of 25 mN. This load corresponds to an indentation depth of approximately 500 nm, about half of the total film thickness. The depth was carefully selected to minimize substrate influence, which becomes significant at higher loads; for instance, at 100 mN, the penetration depth reached ~1400 nm, fully penetrating the films and extending about 200-500 nm into the MgO substrate. It is important to note that the residual stresses in the film significantly influence the indentation based toughness measurements [37]. The superlattices exhibit substantial compressive stress, as identified by RSMs (Table 1), which is beneficial for toughness and, in this case, comparable across the samples, allowing for direct comparison.

In this technique, radial crack lengths at a certain load are typically used to quantify the fracture toughness. For sharp cube-corner tips, radial cracking in conventional ceramics typically initiates below 15 mN [38]. However, as seen in Figure 6, and further in supplementary information section 3, no radial fractures forms at 25 mN, or even at 100 mN, unlike in thicker $Mo_{1-x}Ta_xN_y$ coatings [39], where higher loads could induce cracking. Instead, qualitative indicators such as material pile-up, tangential fractures, and possible spallation are used to assess the toughness [40].

Indents in $HfN_{1.33}$ show no sign of cracking but exhibit significant material pile-up, indicative of substantial plastic flow, consistent with its ductile response in micropillar compression in Figure 5b. The presence of pile-up without cracking suggests high toughness. Similar behavior has been reported in tough understoichiometric systems such as $VN_y$ [41], $V_{0.5}Mo_{0.5}N_y$ [42], [43] and $Ti_{1-x}W_xN_y$ at high W-content [44], where toughness is enhanced by high vacancy concentration. In contrast, the indents in $Hf_{0.76}Al_{0.24}N_{1.15}$ show very limited pile up and instead multiple tangential fractures and film delamination/spallation, indicative of brittle behavior and low toughness. The differences are also evident in the load-depth curves: $HfN_{1.33}$ shows smooth loading/unloading, whereas $Hf_{0.76}Al_{0.24}N_{1.15}$ displays multiple pop-ins, reflecting discrete fracture events (Supplementary Section 3).

The superlattices (Figure 6f-h) exhibit a mixed response, combining features of both parent materials. Moderate pile-up alongside a limited number of tangential cracks and localized spallation suggests an intermediate toughness, higher than the brittle $Hf_{0.76}Al_{0.24}N_{1.15}$, but lower than the ductile $HfN_{1.33}$. Unlike the micropillar compression results (Figure 5), no significant differences between the various superlattice configurations are observed by cube corner indentation. Load-depth curves for the superlattices show fewer and smaller pop-in events compared to $Hf_{0.76}Al_{0.24}N_{1.15}$, further supporting improved toughness. Thus, the superlattice films exhibit the much sought after combination of high hardness and improved toughness thanks to superlattice effects and multilayer toughening.

# 4 Conclusion

We have demonstrated the successful growth of high-quality, three-fold superstructured HfN$_{1.33}$ / Hf$_{0.76}$Al$_{0.24}$N$_{1.15}$ single-crystal superlattices on MgO(001) substrates, exhibiting two distinct 3D checkerboard superstructures with characteristic periods of ~7.5 Å (HfN$_{1.33}$) and ~12.5 Å (Hf$_{0.76}$Al$_{0.24}$N$_{1.15}$) alongside tunable superlattice periodicities of ~20 nm, ~10 nm, and ~6 nm. Structural characterization using XRD, RSMs and STEM confirmed the epitaxial quality, low mosaicity, and well-defined interfaces, with RSMs also revealing a small lateral coherence length due to thin columnar grains. The relaxed lattice parameter and in-plane compressive strain (~4.493 Å and -1.1 %, respectively), closely matched the average values of the monolithic reference films. An unidentified minority phase, exhibiting signs of a four-fold symmetry, was consistently observed in all samples and is strongly correlated with the single crystal peaks.

Mechanical testing via nanoindentation, micropillar compression, and cube-corner indentation revealed that the superlattices retained the high hardness (~37 GPa) of Hf$_{0.76}$Al$_{0.24}$N$_{1.15}$, independent of the superlattice period. In contrast, the strength under uniaxial compression (~7.0 GPa) followed the rule-of-mixture relative to the monolithic constituents. Micropillar compression revealed a more distributed and recoverable fracture behavior in the superlattices, dominated by less severe cracks along the {110}<110> system (same as Hf$_{0.76}$Al$_{0.24}$N$_{1.15}$), suggesting that the coherent interfaces effectively deflect and arrest crack propagation. Crack fragmentation increased with decreasing bilayer thickness, with the most extensive crack dispersion observed in the $\Lambda = 6$ nm superlattice, indicating enhanced fracture toughness. This trend was corroborated by cube-corner indentation, where the superlattices exhibited intermediate toughness between the brittle Hf$_{0.76}$Al$_{0.24}$N$_{1.15}$ the tough and plastically deformable HfN$_{1.33}$.

While the exceptional plasticity and toughness of overstoichiometric HfN$_{1.33}$ remain unmatched, the ability of the HfN$_{1.33}$ / Hf$_{0.76}$Al$_{0.24}$N$_{1.15}$ superlattices to simultaneously retain high hardness and enhance toughness highlights the effectiveness of this multilayer design strategy. These findings underscore the potential of combining hard yet brittle phases like cubic HfAlN with plastically deformable nitride architectures to engineer next-generation coatings with superior damage tolerance. Future designs that further optimize interface chemistry, coherency strain, and superstructure periodicity may unlock even greater synergy between strength, plasticity, and toughness in refractory ceramic systems.

# 5 CRediT Author Statement

**Marcus Lorentzon:** Conceptualization, Investigation, Validation, Data Curation, Visualization, Formal analysis, Writing – Original Draft, Writing – Review & Editing. **Rainer Hahn:** Investigation, Formal analysis, Writing – Review & Editing. **Justinas Palisaitis:** Investigation. **Helmut Reidl:** Funding acquisition. **Lars Hultman:** Supervision, Funding acquisition, Formal analysis, Writing – Review & Editing. **Jens Birch:** Supervision, Funding acquisition, Writing – Review & Editing. **Naureen Ghafoor:** Supervision, Investigation, Funding acquisition, Writing – Original Draft, Writing – Review & Editing.

# 6 Conflict of Interest

The authors declare no conflict of interest

# 7 Acknowledgement

The authors gratefully acknowledge the financial support from the Swedish Research Council, VR, 2018-05190_VR and the Swedish Government Strategic Research Area in Materials Science on Advanced Functional Materials (AFM) at Linköping University (Faculty Grant SFO Mat LiU No. 2009 00971). Swedish Research Council and Swedish Foundation for Strategic Research are acknowledged for access to ARTEMI, the Swedish National Infrastructure in Advanced Electron Microscopy (2021-00171 and RIF21-0026). The financial support by the Austrian Federal Ministry for Digital and Economic Affairs, the National Foundation for Research, Technology and Development, and the Christian Doppler Research Association is gratefully acknowledged (Christian Doppler Laboratory, Surface Engineering of High- performance Components).